# Rapid Application Evolution and Integration Through Document Metamorphosis

Paul M. Aoki        Ian E. Smith        James D. Thornton

Xerox Palo Alto Research Center
3333 Coyote Hill Road
Palo Alto, CA 94304-1314 USA

**Abstract**

The Harland document management system implements a data model in which document (object) structure can be altered by mixin-style multiple inheritance at any time. This kind of structural fluidity has long been supported by knowledge-base management systems, but its use has primarily been in support of reasoning and inference. In this paper, we report our experiences building and supporting several non-trivial applications on top of this data model. Based on these experiences, we argue that structural fluidity is convenient for data-intensive applications other than knowledge-base management. Specifically, we suggest that this flexible data model is a natural fit for the decoupled programming methodology that arises naturally when using enterprise component frameworks.

## 1   Introduction

Modern application frameworks provide a wide variety of facilities that both simplify deployment and enhance the scalability of data-intensive applications. To exploit these frameworks and their services, programmers break applications into modular sub-programs (e.g., "servlets") and reusable components. These sub-programs generally communicate through application program interfaces (APIs) provided by the framework.

Application frameworks generally include database access interfaces and message queue support. For example, the Java™ 2 Platform Enterprise Edition [32] (J2EE) provides the JDBC™ and Enterprise JavaBeans™ (EJB) APIs for database access and the Java™ Message Service (JMS) API for message queuing.[1] Applications can use reliable storage and queuing as a means of *decoupling* data producers from data consumers [37]. By adding a layer of indirection, decoupling often contributes to both modularity and scalability.

Frameworks like J2EE, however, provide quite distinct facilities for the communication of messages and the general management of persistent data for an application, despite the fact that messages will often concern and revolve around significant persistent data. They do not provide the flexibility to allow persistent records of application objects themselves to be the communication vehicles within decoupled applications. In fact, many frameworks do not make any decoupling of parts of applications convenient because they do not provide useful structuring facilities for (e.g.) message headers and metadata. This is true even of tuplespace models [5], which take unification of data management and communication as a primary goal.

The Bantam API, developed at Xerox PARC, defines a data model in which the structure of data for persistent "objects" can be altered by *metamorphosis* – mixin-style multiple inheritance [33] – at any time. With this flexibility, an object that is part of the persistent state of the application can also be the vehicle for communication between producers and consumers, augmented with necessary data in-place as an alternative to message headers. At the same time, parts of the application can evolve in a decoupled way by reusing data objects in non-interfering ways. This kind of structural fluidity has long been supported by knowledge-base systems, but its use has primarily been in support of reasoning and inference (e.g., the dynamic construction of ontologies [10]).

---

[1] Throughout this paper, we will primarily use Java standards and APIs as examples when we refer to various types of software infrastructure. This is due to their wide exposure and is not intended as a slight to other systems, particularly those that predate their Java equivalents (e.g., IBM MQSeries™ or Oracle Advanced Queuing Option™ in the case of message queue systems).

With the Bantam API, we have been able to gain experience with structural fluidity for several non-trivial applications outside the context of traditional knowledge-base management. We suggest that this flexible data model is a natural fit for the decoupled programming methodology that arises naturally when using application frameworks. Bantam promotes a program structure that "feels" like object persistence, but encourages queue-like decoupling

The paper is organized as follows. In Section 2, we give a brief overview of the data model underlying Bantam and the current implementation of that data model in the Harland system. In Section 3, we describe how we documented our experiences – both the systems we studied and the means by which we gathered the information. Section 4 provides an overview of a Bantam application as context for Section 5, which describes the specific lessons we learned. In the final two sections, we place Bantam and Harland in their context in the academic literature and then summarize our findings.

## 2 Current system

The rest of the paper requires some understanding of both Bantam and Harland. Since neither has appeared in the literature to date, we provide a brief overview.[2] In this section, we first describe the data model supported by the Bantam API. We then sketch the current implementation of the Bantam API in the Harland document management system.

### 2.1 Data model

The current model evolved from a simple property store. Property stores are an old idea, dating back at least to logic databases based on (physical) *k*-ary relations [20]. Our notion of a property is more complex than that of such systems.

The Bantam data model features three primary elements: *documents*, *properties* and *schemas*. Documents roughly correspond to persistent objects and have system-generated identifiers. Properties are named data values associated with a document. Schemas are named groups of properties and impose constraints (e.g., type and arity) on their member properties. A Bantam data space consists of a set of documents and their properties, with schemas defining significant structures that documents can assume.

Documents are primarily persistent containers for properties. However, the Bantam model also includes *collection documents* (documents containing other documents as members) and *content documents* (documents having file contents separate from their properties and accessible through a read/write API). A document may have an arbitrary number of properties, each one uniquely identified by a simple text name.

A document property has a name and one or more values. A value is typically a simple data value, such as an integer or a string.[3] Properties with multiple values have bag semantics, i.e., there may be duplicates and no consistent ordering is guaranteed. Individual property values may be added to or removed from documents at any time as long as schema constraints are not violated.

A schema is a named group of properties with constraint definitions. A schema defines a regular structure that individual documents may have and typically corresponds to a "role" that a document object may play within an application. The constraints on a property specify the data type of its values and restrict the number of values. For example, the abstraction of a "to-do list item" might be defined by a schema with four properties: `Subject`, a single text string; `Received` and `Deadline`, single date fields; and `Categories`, an optional text string property that may have many values. Schemas may overlap (contain the same property or properties), but definition of inconsistent schemas is disallowed.

A document with properties satisfying all the constraints of a given schema is said to *conform* to that schema. A schema may also be *enforced* on a conforming document, which causes the system to reject any modifications that would change the document to become non-conforming. A document does not have to have a single or even primary schema; it may have any number of schemas enforced, or none at all. Schemas are enforced on individual documents and may be enforced or unenforced at any time as long as there are no conflicts. In addition, documents may have any number of properties that are not part of any enforced schemas, again as long as there are no conflicts with the enforced constraints.

Queries may be used to retrieve sets of documents based on the values of their properties and/or the schemas enforced upon them. A variety of common query operators for property values are supported and composite queries may be built using AND, OR, and NOT but the model provides no support for joins. There are some additional query operators to address collection membership, content indexing, and single vs. multiple values.

---

[2] Additional, detailed information about the Bantam API and the Harland system may be obtained from the system documentation [34] contained in the public Harland software distribution. As of this writing, this distribution can be obtained from the Xerox PARC Web site, http://www.parc.xerox.com/harland/.

[3] The current implementation, written in Java, does permit arbitrary `Serializable` objects as values.

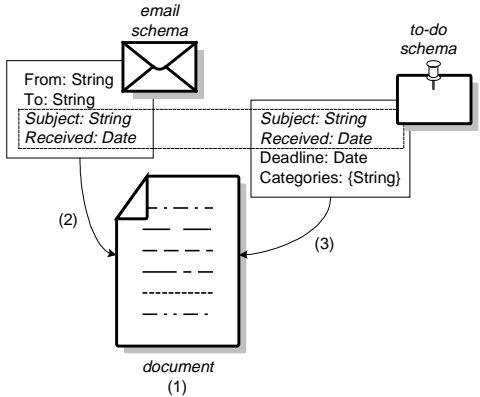

**Figure 1: Bantam schema usage.**

The support for structural fluidity in the Bantam data model derives from the fact that properties can change and schemas can be enforced and unenforced throughout the execution of applications. The mixin-style multiple inheritance of document structure derives from the fact that documents may have a mix of schemas enforced or property groupings that have yet to be codified by schema at all.

We now give an example of schemas usage. Figure 1 shows a base document with textual content. The document enters a user application *A* as an email message and is added to the repository (step (1)). Because of its source, application *A* might choose to enforce the "email" schema after setting the requisite property values (step (2)). Later, an application *B* using the same repository might detect the addition of the document, classify it as a request for action, and choose to enforce the "to-do" schema on the same document (step (3)). Observe that some of the to-do schema properties overlap with the email schema properties; in this case, the property values are fully shared between applications *A* and *B*.

The data model described above evolved through experience with building and using two earlier property store systems. The first of these, Presto [8], featured a simple data model that was very close to the present Bantam model except that it did not include schemas or any other form of enforceable constraint. The Placeless Documents system [9] followed Presto and featured a considerably more complex data model in which properties existed in hierarchies on documents, multiple instances of properties were supported with distinct sub-properties, and every document could have both shared properties and personal properties for each system user. As with Presto, Placeless Documents did not intrinsically support schemas or constraints on properties. Both Presto and Placeless Documents were created to explore novel approaches to document management by end-users. The Placeless Documents system, however, was also used as infrastructure for the development of a number of internal applications. Through their experience writing such applications, programmers recognized a need for documented structures and enforceable constraints. Bantam-style schemas first appeared in a support library for Placeless Documents. Thus Harland was grounded from the beginning in software development experience.

## 2.2 Implementation

Harland is our current implementation of the Bantam API. The entire system is written in Java. The current release contains about 26,000 lines of code (26 KLOC).

Figure 2 provides an abstract view of the Harland internal architecture. Harland is organized into five main sections, indicated by the shaded backgrounds in the figure. The first two sections implement a middleware query processor and a content management interface, respectively. A third section manages the organization of properties in a relational store as well as the associated metadata. The final two sections provide "glue" between Harland and the Bantam API, and between Harland and its underlying storage mechanisms. We now discuss each of these sections in turn:

*Query processor*. Harland includes a relatively generic middleware query processing architecture. Query rewrite and optimization produces a graph-structured query plan from the application's query. A query processor interprets the query plan. Since Harland is built on top of a relational store, the query processor constructs SQL queries, issues them to the backend DBMS and assembles the rows into an internal document representation (IDoc).

IDocs are inserted into a cache as they are constructed. At the Bantam level, the application sees only a handle to the cached IDoc and never obtains a reference to the IDoc itself, which simplifies cache eviction. Memory is conserved by incremental (demand-driven) materialization of SQL query results and by soft cache size limits.

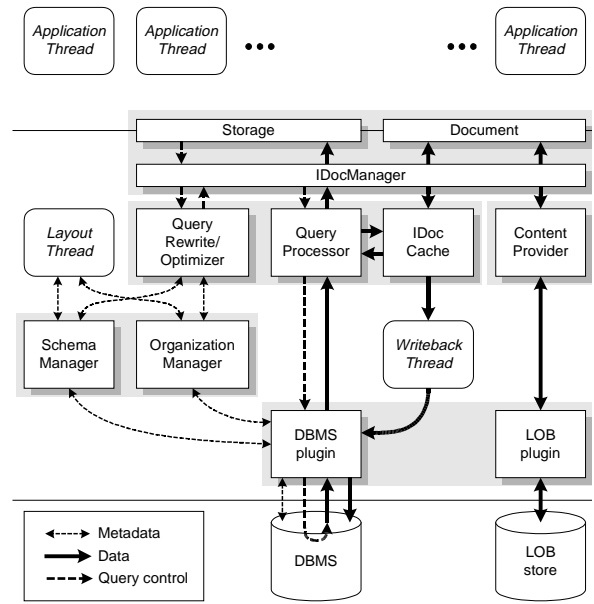

**Figure 2: Harland internal architecture**.

The system is internally multithreaded as well as thread-safe at the API level. For example, an application thread can be accessing the properties of a cached document while a second application thread is issuing a query and the internal writeback thread cleans the document cache "in the background." Internally, document-level lock management is centralized in the cache interface.

*Content management.* Harland's content management section is similarly generic. This section exports a simple interface similar to "large object" (LOB) interfaces. The LOB (content) store need not be the same as the property store.

*Property storage management.* The most interesting (and currently least complete) part of the system deals with the organization of the properties into rows (the organization manager), the organization of properties into schemas (the schema manager) and the metadata needed to manage changes to both.

An important idea behind the design of Harland is that the formal expression of application data structures through schemas provides opportunities for automated optimization. Harland presently uses schemas as implicit hints about data access patterns for automatic prefetching. When an application first accesses a document, the properties are normally not all materialized into the cache immediately, to avoid loading the cache with data that may go unused. Instead, Harland loads (and stores) properties in schema groups whenever possible, so if one property in a schema is required, all the properties in that schema will be loaded. Harland prefetches aggressively to reduce round trips to the backend DBMS. For example, the translation of application queries to SQL is arranged so that the result stream from the DBMS query includes all properties contained in any schema touched by the original query.

Properties are stored in cache `IDocs` in groups called internal slices (`ISlices`) that are intended to be minimal units of co-retrieval. Part of the role of the organization manager is to define `ISlices`. Schema groupings cannot be used directly because they are subject to constant change. Presently the schema manager uses a simple fixed allocation strategy that is heavily influenced by schemas but also somewhat dependent on order of accesses. More sophisticated dynamic strategies are possible but are a matter for future work.

Schemas can also serve as hints for decomposing properties into tables, which would also be managed by the organization manager. For example, all the single-valued properties of a schema could be laid out as columns of a table. Given the highly dynamic nature of the data model, the decomposition decisions and the corresponding query translations could become quite complex. Although anticipated in the design of Harland, such sophisticated approaches to backend data management have not yet been implemented. Presently, Harland uses an organization with one row per property.

The schema manager maintains the persistent record of all enforced schemas, detects conflicts, and keeps track of which schemas are enforced on which documents. The persistent records of schemas and enforcement are currently maintained in the same DBMS as the property data but in separate tables.

*Bantam API support.* Bantam's data structures and interfaces were designed to be intuitive and object-oriented. However, Harland's internal document structure is designed with efficient construction and caching in mind. Similarly, Harland's internal query structure makes assumptions about Harland's architecture that have nothing to do with Bantam per se. Translations between Bantam and Harland data structures occur in a layer that "sits above" the rest of Harland.

*Storage interface support.* Implementation-specific aspects of both the property store and the LOB store are factored into extension interfaces. Harland currently supports Oracle 8i and PostgreSQL[4] as property stores, with support for Hypersonic SQL[5] under development. Either the DBMS used for the property store or the local file system may be used to store document content. DBMS access is through the JDBC™ API.

## 3 Method

Bantam defines a programming interface, and Harland is an implementation of that interface; both are exposed to programmers rather than end-users. Consequently, the experiences reported in this paper are distilled from exercises in application programming. However, the applications described here were not academic software in the sense of software written solely by the authors for their own use. Bantam and Harland have been used by a variety of programmers, both inside and outside of PARC. All but one of the applications described here were deployed for use, one for nearly two years.

In this section, we discuss how we collected the experiences reported in Section 5. We first (briefly) describe the applications in question. We then describe our data collection and analysis methods.

### 3.1 Applications studied

We describe applications built on top of the Bantam API. These applications have been written over the last two years and, as previously mentioned, the API and its implementations have evolved.

The following non-commercial systems have been built on top of Bantam within PARC, some of which were

---

[4] http://www.postgresql.org/

[5] http://hsql.oron.ch/

deployed within our organization for a significant period of time:

*Dealer*. The first application was a Web-based internal scheduling application. The system consisted of 2 KLOC, written in a few days by one developer and in continuous use for nearly two years.

*Raton Laveur* (version 1). The next application was a prototype of a personal information management (PIM) system [2]. The first major version of Raton Laveur itself went through an iterative design process over the course of seven months. The final result was 12 KLOC, again written by one developer.

*Placeless Portal*. Members of the Placeless Documents project built a framework for rapid development of workgroup information portals. This framework consisted of 15 KLOC, written by four developers over a period of three months.

*Raton Laveur* (version 2). The second major version of Raton Laveur used electronic mail as a channel for tracking personal activity. A variety of end-user, message-driven applications – calendar management, sales lead management, etc. – were subsumed into a common infrastructure. The various applications shared a substantial amount of state. The project consisted of over 60 KLOC and six developers over the course of six months.

Another application, *RNC*, is being written using Harland as a persistent store. This project has been underway for about three months with four active developers. However, it is not yet complete, so it will not be discussed further.

Two commercial software projects within Xerox are evaluating Harland for adoption as a repository component. As part of this evaluation, Xerox programmers have written software to "bridge" between the internal object models of the existing application software and Harland's document model. The two applications are:

*QE2*. This is a job-management system, originally developed at Xerox PARC, specialized for the task of printing and shipping technical documentation on demand.

*DocuShare™*. DocuShare™ is a document management system that was originally developed at the Xerox Webster Research Center and is now a commercial product [26].[6] The system provides a customizable Web-based user interface as well as underlying facilities that are very similar to those of WebDAV [12]; however, unlike WebDAV, documents are strongly typed, and these types strictly determine the documents' properties.

### 3.2 Data collection and analysis

It would take an large amount of quantitative data to draw conclusions in a qualitative (e.g., statistical) sense. In this subsection, we mean "data" in the sense of, e.g., qualitative sociology – that of evidence generally supporting a hypothesis.

The lessons described here were obtained by various means. Some of them are taken from first-hand experience, i.e., from involvement in the construction of the software. Some of them are derived from inspection of software produced by other programmers. Finally, some of the information presented here is based on interviews with application developers.

We performed informal, "semi-structured" interviews – i.e., based on a known initial agenda but not fixed to a standard questionnaire. This form of interview is frequently used in iterative design [24] and is particularly appropriate for eliciting comments on issues other than those that have been preconceived as relevant [21]. Questions generally increased in specificity from the very general ("Have you written a database application before?") to the merely broad ("Can you compare Harland's ease of use to that of JDBC?") and the fairly narrow ("Did you define your schemas in a separate data design step? If so, did you find it necessary to add schemas after that design step?"). We interviewed all of the application developers who were available to us (i.e., who were still employed by Xerox). This constituted a group of ten interviewees. Group sizes of this order are generally too small for meaningful quantitative analysis but are quite common in ethnographic studies and system design evaluations.

Once we had gathered our notes, we performed a standard topic clustering [21] to identify common themes. These themes are what have been summarized in the next two sections.

## 4 Inside a Bantam application

It is difficult to explain some of the general experiences of Bantam users without explaining how a Bantam application might be structured. In this section, we give a brief introduction to a sample Bantam application. We have formulated this discussion in general terms, based on our own experiences, code examination and some interview content.

Most of our applications have been built in an application server environment. In particular, much of our experience has been with applications built using the commercial Orion Application Server from Ironflare AB.[7] One application was built using the freely-available Tomcat platform.[8] Note that the use of an application server – even a servlet engine like Tomcat – does not mean that the applications are necessarily Web-based. For example, as described in Section 3.1, Raton Laveur was structured around electronic mail and other types of external messaging infrastructure.

---

[6] http://docushare.xerox.com/

[7] http://www.orionserver.com/

[8] http://jakarta.apache.org/tomcat/

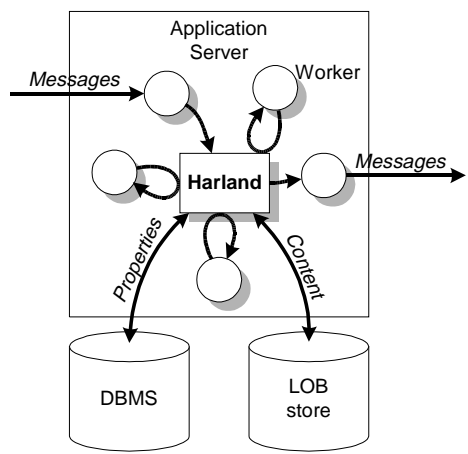

**Figure 3: Bantam application example.**

Figure 3 shows how a message-driven application is constructed using Harland as an internal work-queue system. Specifically, it suggests a design pattern involving the use of a data structure that is really more like a tuplespace [5] than a queue. Here, *worker threads* continually query for documents whose properties and/or enforced schemas match their input criteria. These documents are processed and then made available for other workers. Documents are never removed from the work-queue, since the work-queue is simply the Harland repository. Instead, workers explicitly synchronize by setting new document properties or modifying the values of existing properties and enforcing a schema.

Documents in a Bantam application correspond to different kinds of modeled entities. Some "documents" correspond to what we normally think of as electronic documents, e.g., incoming and outgoing email messages and their attachments, Web pages produced as part of a user interface, etc. Other "documents" correspond to what we often think of as application objects. For example, users are modeled using documents; document properties contain passwords, preferences, and other user-related application state. Finally, some documents correspond to nascent or partial results of some computational process. For example, consider the assembly of an automated email reply. Different portions (e.g., attachments or enclosures) will be processed separately and then assembled into the final result that is sent to a recipient. While each portion is being processed, it may have an existence separate from the final email message. This might make sense, for example, if it had been reused from a previous reply message.

Putting it all together, we see that the programming model represented by Figure 3 is not quite like that of a conventional database-backed application, a tuplespace application, or a queue-structured application. Unlike the database application, Bantam documents serve as a basis for internal application events and coordination; for example, the addition of an "email attachment" schema to a document might signal that the attachment must be encoded for email transmission. Unlike tuples in a tuplespace application and messages in a queue application, Bantam documents are not inserted into an infrastructure by a sender and consumed by a recipient; instead they are persistent records around which coordination occurs.

## 5 Summary of experiences

In this section, we discuss a wide variety of lessons learned from building Bantam applications. These vary somewhat in their level of abstraction. First, we describe some general advantages of Bantam over alternatives as described by the application programmers. Second, we turn to a more specific description of ways in which it (qualitatively) appeared to speed application development. Third, we discuss how Bantam mechanisms appeared to simplify the integration of applications and fragments of applications within a large system. Finally, we share some difficulties introduced by the Harland implementation.

### 5.1 General applicability

Our developers felt that the Bantam model provides a combination of flexibility, structural support, and conceptual accessibility that effectively supports development of persistence features for a variety of applications. The specific advantages they described fell into three general categories: the benefits of dynamic object structure, the (potential) benefits of strong physical data independence, and the benefits of providing a data model other than that provided by the programming language.

*Dynamic data structure instances*. Harland is well suited to the development of any application in which document structure can vary over the lifespan of the document. For example, DocuShare™ manages a variety of persistent objects of direct interest to users, such as documents and calendars. An administrator can customize the definitions of these user-visible objects at runtime, so the repository must have a corresponding degree of flexibility.[9]

This flexibility is consistent with the use of property stores in general. (For this reason, electronic catalogs often use data representations similar to property stores [16].) However, simple property stores do not provide enough structure to satisfy programmers working in strongly typed implementation languages. Bantam better supports the practice of programming by providing the mechanism of schemas, which codify decisions about how to use the property store in a way that is automatically enforceable. This provides developers with reliable documentation of the regular structures that exist. It also eliminates the need for much verification code in the application without giving up the assurance that errors will be automatically detected early. These are exactly the benefits associated with strong type-checking, applied to the context of a flexible property store.

---

[9] Many data models (e.g., those defined by common Internet standards such as WebDAV [12], LDAP [36], RDF [19], and XML with XML-Schema or DTDs) have very flexible structuring primitives. However, the equivalent of schema definition and enforcement is typically performed using mechanisms such as configuration files. Most repositories do not permit on-the-fly structural modifications like those described here. However, see Section 6.

*Potential for storage optimization.* The specification of structural regularities in terms accessible to the system also opens great opportunities for optimization of storage layout and retrieval using a relational DBMS, as introduced earlier in section 2.2.

*Loose language binding.* While a close conceptual match to the implementation language is certainly an advantage, as discussed in the next section, one can also argue that the Harland distinction between the persistence mechanisms and the basic programming language mechanisms is valuable. It may be helpful in general to consider persistence questions, such as which data items should be preserved, separately from the class structure of an application implementation.

One advantage of such a separation is that it permits structural arrangements for persistent data that cannot be directly represented in the class structure of the implementation language, such as multiple-inheritance structures with a Java implementation.

Another possible advantage is that the independent evolutions of parts of a system, which are the focus of section 5.3, are less likely to conflict if they do not necessarily involve altering large shared implementation class structures.

Finally, we note that even with a persistence middleware layer closely matched to the implementation language, it is still desirable to think of the persistent data organization separately from the code because there is a significant difference between working copies of data in main memory and persistent data on disk. Code may also need to evolve without requiring all stored data to be manipulated. We think that this justification is analogous to the argument that the design of distributed systems should not attempt to hide the distinction between local and remote objects, because that distinction is fundamental and important [18].

## 5.2 Rapid development

A common feeling among the developers was that Bantam "kept simple things simple." They shared some more specific observations as well. These can be summarized as follows: interface simplicity, appropriate abstractions, and the ability to defer (or even eliminate) some aspects of persistent object design and inter-developer coordination.

*Simplicity.* Because the Bantam model is so simple, application programmers uniformly found it to be more accessible than the standard alternatives. For example, issuing a JDBC query requires a large amount of setup by comparison. JDBC has many abstractions that are due to its SQL-based nature (e.g. `java.sql.Statement`). Programmers preferred Harland because it introduced fewer novel abstractions. J2EE requires the programmer to learn even more abstractions. Adding to the convenience, the Bantam API tightly binds Harland functionality to the Java language, permitting schemas to be defined by stylized class definitions that are legal Java code for example.

Of course, simplicity comes at the price of expressiveness, the effects of which we discuss in Section 5.4.

*Document-centric abstractions.* The Bantam abstractions are closely related to basic application abstractions for document applications and to the object abstraction for general applications. Fitting application models closely is a strength of standard object-oriented systems, whether OODBs or O-R mapping systems, but these do not offer the structural fluidity of Harland and can have considerable complexity.

*Lazy interface design.* Several programmers were also enthusiastic about the ability to defer the precise definition of classes. Note that the existence and purposes of the documents themselves and their inter-document relationships would usually be defined in a conventional "class design" step. These would generally be accompanied by an initial schema definition. However, schemas were often "layered on" during the construction of software. Sometimes, this was because the need for some properties was recognized between producers and consumers of information, the nature of which would not be known until the producer and consumer were implemented concretely.

*Lazy developer coordination.* By hiding data layout and supporting arbitrary combinations of schemas and non-schema properties, Harland permits extensions to be made easily and with localized impact as an application evolves.

One subtle advantage of this simplicity, identified by one programmer, was the elimination of a strongly distinct "database group" within the development team because persistence facilities were so accessible, and the flexibility and physical independence minimizes the need for strong central coordination of all changes.

## 5.3 Application integration

Section 4 describes a general structure that could be naturally constructed using (e.g.) a tuplespace system [5]. Bantam allows programmers to define structure for documents incrementally and dynamically (unlike Linda and TSpaces, which provide unstructured tuples, or JavaSpaces [11], which provides structured Java objects).

It is useful to think of this as a way to allow two pieces of software (a producer and a consumer) to agree on a document "view" without forcing the rest of the software infrastructure to be aware of this view (as would necessarily be the case if documents had a fixed membership in a predefined class).

There are two ways in which we observed application integration occurring with the Bantam model. (Most of our experience in this area is based on the second version of Raton Laveur.) We taxonomize these by how they used schemas:

*Shared documents with disjoint schemas.* Individual objects may become part of distinct applications that are highly independent because each application can enforce its own schema independently of all the rest. For example, an object modeling an email message could become a to-do list item for another application. Assuming the use of simple conventions to avoid accidental property name collisions, the email and to-do list applications could operate quite independently, sharing nothing beyond a known object identifier.

Coordination between such applications, when necessary, occurred through shared schemas. For example, a separate schema might be defined to provide a synchronization token of some kind. In fact, sometimes a schema was defined with no properties at all and enforced purely for the purpose of synchronization.

*Shared documents with overlapping schemas.* Applications may selectively interoperate with the reliability of enforced structure at their interface. This would be the case with the producer/consumer example mentioned earlier. The schemas in this situation serve a role analogous to a formal interface definition in an architecture like CORBA [25]. Harland's support for multiple-inheritance through multiple, and possibly overlapping, schemas allow applications (or pieces of applications) to store whatever data they require and coordinate on only a subset of that data.

In fact, users of Harland for the second version of Raton Laveur encountered cases where multiple consumer components operated on the same objects without explicit coordination between programmers, made possible because the key interface elements were codified in schemas.

We did not observe s*hared documents with subset schema.* In general, schemas are small enough that the explicit construction of subset schemas (equivalent to simple projection views) was not necessary.

### 5.4 Gaining non-trivial experience

The problem with trying to support applications that will actually be deployed is that the implementors need to think about production requirements. The software is fairly stable, and we gained many practical benefits (hot backup, etc.) just from using a relational database as the persistent store. However, there are several important features missing when compared to (e.g.) a full J2EE implementation. Some were "simple" in the sense of being implementation holes; most notably, Harland does not currently have a useful range of transactional semantics. Aside from these basic problems, programmers had two major categories of complaints. We summarize these as a lack of scalability and a lack of now-common application services.

*Lack of scalability*. Harland is a research prototype and is generally less scalable than a J2EE implementation. Because Bantam defines its own non-relational data model, we necessarily have to do additional query processing and structural mapping to construct Bantam documents, which takes up CPU. We try to reduce operation latency by caching documents and avoiding frequent connection "context switching," but this ties up memory and other resources, further reducing throughput. Finally, the current prototype cannot easily support large-scale parallelism techniques, such as spreading computation across clusters of machines.

*Lack of application services*. Harland is not a component of any existing entire application framework. A J2EE application (to use an example of such a framework) is perceived as "automatically" providing a host of services; the use of Harland in place of any of its component technologies (e.g., EJB) means that none of the services provided by that component are available.

These complaints (which, it should be noted, are complaints about the Harland system rather than the Bantam API) are quite serious. They have, for example, resulted in the third version of Raton Laveur being written against J2EE/EJB.

## 6 Related work

We first discuss the relationship of Bantam and Harland to other proposals for providing flexible object structure. We then relate the programming style and structure described in Sections 4 and 5 to previous work.

### 6.1 Dynamic object structure – primitives and implementation

A wide variety of systems have attempted to provide flexible object primitives. The general goal – shared with Bantam – has been to address the fact that the data structures (object design) of an application must be able to cope with the addition of new types (evolution) in a way that is intuitive for programmers.

Many, if not most, of the key developments in flexible persistent data representation came from early work in knowledge representation (KR) and knowledge-base management systems (KBMS). The trick of layering arbitrary dynamic structures (e.g., frames [10]) on top of simple, fixed 2- or 3-tuples ("vertical schemas") works as well today in PERK [17] as it did in 1967 [20]. (Interestingly, vertical schemas have recently become popular as an implementation technique for electronic catalog applications [16] for almost exactly the same reasons as those given by KR practitioners.) KBMS implementers have explored both manual (e.g., KEEconnection [1]) and semi-automatic (e.g., KRISYS [7]) decomposition of objects into tables.

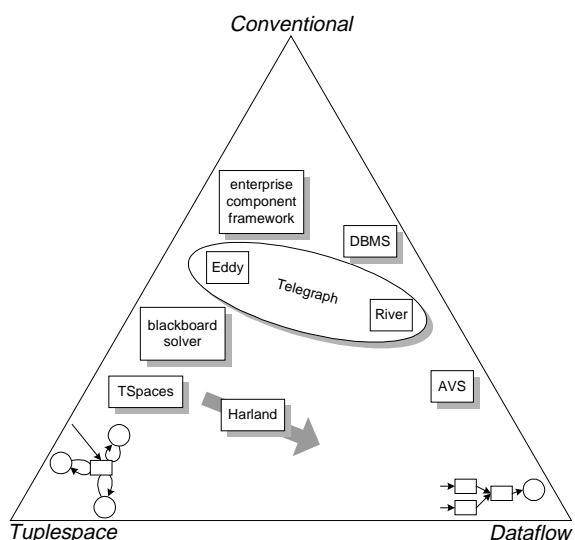

**Figure 4: Models for data-intensive programs.**

Structural flexibility in object databases has been influenced by the KR/KBMS work as well as by later developments in programming languages [33]. Most object management systems use some kind of simple, fixed object decomposition strategy (e.g., VISION [28], clovers [30], roles [13], aspects [27], SaveMe [3]), perhaps with manual input to override the default strategy (e.g., JavaBlend™ [31]). A few systems have attempted to provide semi-automatic algorithms for object decomposition (notably Iris [22]).

Repositories for various kinds of typed information often permit very flexible multiple inheritance schemes. WebDAV [12], RDF [19] and LDAP [36] have similar object models in this regard, and the vertical schema implementations of DocuShare [26], RDFdb[10] and IBM Enterprise Directory [29] reflect this. Microsoft Meta Data Services (formerly Microsoft Repository) uses a simple object decomposition strategy with manual overrides [4]. Semistructured data repositories share the goal of "flexible" structure, but not the specific manner of flexibility.

We believe that the systems research goal of Harland – the use of schemas as automatic storage layout hints – is unique.

### 6.2 Object structure vs. application structure

The discussion of Sections 4 and 5 on program structure give rise to a taxonomic question. Consider the three general approaches of conventional (imperative control-flow) programming, tuplespace programming (e.g.,

---
[10] http://www.guha.com/rdfdb/

Tspaces [37]) and dataflow programming (e.g., AVS [35]). Organizing the system design space along these (not entirely orthogonal!) axes, one can begin to place systems within this space. For example, most application framework programs end up in a fairly conventional style. The internal structure of conventional database engines is usually (at least partially) in dataflow; the internal structure of the Telegraph DBMS is partly dataflow (River) and partly like a tuplespace (Eddy) [14]. Figure 4 illustrates this design space graphically.

We believe that the use of Bantam (as implemented in Harland) leads to a programming style that is more explicitly a hybrid of the dataflow and tuplespace styles. Documents act like messages, but they also act like persistent objects; the "message-like" behavior comes about as a result of changes in the object structure.

## 7 Conclusions

We have introduced the Bantam model of structural fluidity through mixin-style multiple inheritance and the Harland prototype implementation. Our experiences with application development using Harland have suggested that there are notable features of this data model for rapid application evolution and integration that commend it for consideration outside the traditional domain of knowledge-base management. The combination of flexibility and assurance of schema enforcement add value for these applications beyond that of many closely related technologies like tuplespaces. We have also found that typical limitations of research software have had the common effect of restricting our ability to gain non-trivial experience with Harland.

As discussed in section 2.2, there is considerably more work to be done in the area of property storage management, particularly concerning the decomposition of properties into tables. Being in the nature of an automatic optimization problem, much more intensive study of the access patterns of particular applications running on Harland is likely to be necessary.

We have delivered two general software releases of Harland. This software is being provided to interested parties with varying degrees of engagement:

First, as we have mentioned, Harland has been provided to groups within Xerox business units. Harland is also in active use by research projects within PARC.

Second, we have initiated cooperative research engagements with selected university research groups. For example, we have funded work at the Group in User Interface Research (GUIR) at the University of California, Berkeley. GUIR co-developed *NotePals* [6], a system that synchronizes, stores and summarizes collections of note-sized documents. The NotePals project has already begun exploring property-based note-sharing [15]. Similar efforts have been undertaken with the Everyday Computing group at Georgia Tech, since this group has similar experience in building collaborative systems (e.g., Flatland [23]) on top of property stores.

Finally, as mentioned in Section 2, the software and associated documentation are available for public download.

We look forward to learning more about the value of the Bantam model of structural fluidity in various application contexts through these collaborations. We expect that further work on Harland itself will be guided by such input.

## Acknowledgements

Paul Dourish, Keith Edwards, John Lamping and Tom Rodriguez contributed to the development of the Bantam API. Karin Petersen helped to crystallize the application integration concepts described here. Hans Koomen wrote the bridge software between the existing document applications and Harland; we appreciate his time, effort and constructive feedback. Jun Gabayan provided the bulk of the PostgreSQL support. Finally, we thank all our colleagues who used Harland for their application development and graciously agreed to be interviewed.

## References

[1] Abarbanel, R.M. and M.D. Williams, "A Relational Representation for Knowledge Bases," in *Proc. 1st Int'l Conf. on Expert Database Sys.* (Charleston, SC, Apr. 1986), L. Kerschberg (ed.), Benjamin/Cummings, Menlo Park, CA, 1987, 191-206.

[2] Bellotti, V. and I. Smith, "Informing the Design of an Information Management System with Iterative Fieldwork," *Proc. 3rd ACM Conf. on Designing Interactive Sys.*, New York, Aug. 2000, 227-238.

[3] Berchtold, S. *et al.*, "SaveMe: A System for Archiving Electronic Documents Using Messaging Groupware," *Proc. Joint Conf. on Work Activities, Coordination & Collaboration*, San Francisco, CA, Feb. 1999, 167-176.

[4] Bernstein, P.A. *et al.*, "The Microsoft Repository," *Proc. 23rd VLDB Conf.*, Athens, Greece, Aug. 1997, 3-12.

[5] Carriero, N. and D. Gelernter, "Linda in Context," *CACM 32*, 4 (Apr. 1989), 444-458.

[6] Davis, R.C. *et al.*, "NotePals: Light Weight Note Sharing by the Group, for the Group," *Proc. ACM SIGCHI Conf.*, Pittsburgh, PA, May 1999, 338-345.

[7] Deßloch, S. *et al.*, "Advanced Data Processing in KRISYS: Modeling Concepts, Implementation Techniques, and Client/Server Issues," *VLDB J. 7*, 2 (May 1998), 79-95.

[8] Dourish, P. *et al.*, "Presto: An Experimental Architecture for Fluid Interactive Document Spaces," *ACM TOCHI 6*, 2 (June 1999), 133-161.

[9] Dourish, P. *et al.*, "Extending Document Management Systems with User-Specific Active Properties," *ACM TOIS 18*, 2 (Apr. 2000), 140-170.

[10] Fikes, R. and T. Kehler, "The Role of Frame-Based Representation in Reasoning," *CACM 28*, 9 (Sep. 1985), 904-920.

[11] Freeman, E. *et al.*, *JavaSpaces™: Principles, Patterns and Practice*, Addison Wesley, Reading, MA, 1999.

[12] Goland, Y. *et al.*, "HTTP Extensions for Distributed Authoring - WEBDAV," RFC 2518, IETF, Reston, VA, Feb. 1999.

[13] Gottlob, G. *et al.*, "Extending Object-Oriented Systems with Roles," *ACM TOIS 14*, 3 (July 1996), 268-296.

[14] Hellerstein, J.M. *et al.*, "Adaptive Query Processing: Technology in Evolution," *IEEE Data Eng. Bulletin 23*, 2 (June 2000), 7-18.

[15] Huang, J. and J. Michiels, "Exploring Property-Based Document Organization in a Collaborative Note-Sharing System," *Extended Abstracts, ACM SIGCHI Conf.*, the Hague, the Netherlands, Apr. 2000, 327-328.

[16] Jhingran, A., "Moving Up the Food Chain: Supporting E-Commerce Applications on Databases," *SIGMOD Record 29*, 4 (Dec. 2000), 50-54.

[17] Karp, P.D. *et al.*, "A Collaborative Environment for Authoring Large Knowledge Bases," *J. Intelligent Info. Sys. 13*, 3 (Nov.-Dec. 1999), 155-194.

[18] Kendall, S.C. *et al.*, "A Note on Distributed Computing," Sun Microsystems Research TR-94-29, Nov. 1994.

[19] Lassila, O. and R.R. Swick, "Resource Description Framework (RDF) Model and Syntax Specification," W3C, Cambridge, MA, Feb. 1999.

[20] Levien, R.E. and M.E. Maron, "A Computer System for Inference Execution and Data Retrieval," *CACM 10*, 11 (Nov. 1967), 715-721.

[21] Lofland, J. and L.H. Lofland, *Analyzing Social Settings*, Wadsworth, Belmont, CA, 1995.

[22] Lyngbaek, P. and V. Vianu, "Mapping a Semantic Data Model to the Relational Model," *Proc. ACM SIGMOD Conf.*, San Francisco, CA, May 1987, 132-142.

[23] Mynatt, E.D. *et al.*, "Flatland: New Dimensiolns in Office Whiteboards," *Proc. ACM SIGCHI Conf.*, Pittsburgh, PA, May 1999, 346-353.

[24] Newman, W.M. and M.G. Lamming, *Interactive System Design*, Addison Wesley, Reading, MA, 1995.

[25] *Common Object Request Broker: Architecture and Specification, Revision 2.4.2,* Object Management Group Inc., Needham, MA, Feb. 2001. See also http://www.omg.org/technology/documents/formal/corbaiiop.htm

[26] Rein, G.L. *et al.*, "A Case for Document Management Functions on the Web," *CACM 40*, 9 (Sep. 1997), 81-89.

[27] Richardson, J. and P. Schwarz, "Aspects: Extending Objects to Support Multiple, Independent Roles," *Proc. ACM SIGMOD Conf.*, Denver, CO, May 1991, 298-307.

[28] Sciore, E., "Object Specialization," *ACM TOIS 7*, 2 (Apr. 1989), 103-122.

[29] Shi, S.S.B. et al., "An Enterprise Directory Solution with DB2," *IBM Sys. J. 39*, 2 (2000), 360-383.

[30] Stein, L.A. and S.B. Zdonik, "Clovers: The Dynamic Behavior of Types and Instances," *Comp. Sci. & Inf. Mgmt. 1*, 3 (1998), 1-11.

[31] *JavaBlend™ Developer Tools Guide (Version 2.0)*, Part No. 806-0569-10, Sun Microsystems, Mountain View, CA, Nov. 1999.

[32] *Java™ 2 Platform Enterprise Edition Specification, Version 1.2*, Sun Microsystems, Mountain View, CA, Dec. 1999. See also http://java.sun.com/j2ee/.

[33] Taivalsaari, A. "On the Notion of Inheritance," *ACM Comp. Surveys 28*, 3 (Sep. 1996), 438-479.

[34] Thornton, J., "Writing Programs for the Harland Property Store," Xerox PARC, Palo Alto, CA, Oct. 2000. See http://www.parc.xerox.com/harland/.

[35] Upson, C. *et al.*, "The Application Visualization System: A Computational Environment for Scientific Visualization," *IEEE CG&A 9*, 4 (July 1989), 30-42.

[36] Wahl, M. *et al.*, "Lightweight Directory Access Protocol (V3)," RFC 2251, IETF, Reston, VA, Dec. 1997.

[37] Wyckoff, P., *et al.*, "TSpaces," *IBM Sys. J. 37*, 3 (1998), 454-474.